\documentclass[global]{svjour}
\usepackage{graphics}
\begin{document}
\title{Autocatalytic reaction on low-dimensional substrates}

\author{E. Agliari\inst{1}, R. Burioni\inst{1, 2}, D. Cassi \inst{1, 2} \and F.M.
Neri\inst{1}}
\institute{Dipartimento di Fisica, Universit\`a degli Studi di
Parma, Parco Area delle Scienze 7/A, 43100 Parma, Italy \and INFN,
Gruppo Collegato di Parma, Parco Area delle Scienze 7/A, 43100
Parma, Italy}
\date{\today}
%
\maketitle
\begin{abstract}
We discuss a model for the autocatalytic reaction $A+B\rightarrow
2A$ on substrates where the reactants perform a compact exploration
of the space, i.e., on lattices whose spectral dimension $\tilde{d}$
is $< 2$. For finite systems, the total time $\tau$ for the reaction
to end scales according to two different regimes, for high and low
concentrations of reactants. The functional dependence of $\tau$ on
the volume of the substrate and the concentration of reactants is
discussed within a mean-field approximation. Possible applications
are discussed.
\end{abstract}
\section{\label{sec:intro}Introduction}

Diffusion-reaction processes is a long standing problem which
finds a number of applications, especially in physics
\cite{bartumeus}, chemistry and biology \cite{hess}.


Most of the earlier studies focus the attention on a single
particle diffusing in the presence of immobile reactants, while
much less is known about the statistical properties associated
with the diffusion of a set of particles, notwithstanding its
interest. Indeed, multiparticle diffusion problems are difficult
to manage due to the fact that the effects of each single particle
do not combine linearly, even in the noninteracting case
\cite{yuste}. In the last years much effort has been devoted to
the formulation of rigorous many-body treatments of diffusion
controlled reactions especially in low dimension. In fact, while
in high dimensions a mean-field approach provides a good
description, in low dimension local fluctuations are responsible
for significant deviation from mean-field predictions
\cite{toussaint}.

In general, a great deal of recent experimental as well as
theoretical work has been devoted to the study of such
diffusion-reaction processes in {\it restricted geometries}. The
latter expression refers to two different (possibly concurrent)
situations: that of a low dimensionality, and that of a small
spatial scale. In the first case, the spectral dimension $d$
characterizing the diffusive behavior of the reactants on the
substrate is low ($1 < d < 2$), and the substrate underlying the
diffusion-reaction lacks spatial homogeneity. Hence, there are
considered media whose properties are not translationally invariant
and where the reactants perform a ``compact exploration'' of the
space \cite{degennes}. These kinds of structures can lead to a
chemical behavior significantly different from those occurring on
substrates displaying an homogeneous spatial arrangement. This is
the case, for example, of fractal lattices: in the last 20 years, an
extensive literature has been investigating the consequences of a
fractal geometry on the laws of reaction kinetics \cite{blumen}, for
example for the one-species ($A + A \rightarrow
\emptyset$)\cite{zumofen1} and two-species ($A + B \rightarrow
\emptyset$)\cite{zumofen1,lindenberg,zumofen2} annihilation
reactions. In all these systems the role of the generally noninteger
spectral dimension, whose definition will be discussed below, is
stressed, as opposed to the integer euclidean dimension
characterizing homogeneous structures.

But {\it restricted geometry} also refers to a variety of
experimental situations in which these processes occur on spatial
scales too small to allow an infinite volume treatment. The
so-called finite-size corrections to the asymptotic
(infinite-volume) behavior in this case become predominant.
Indeed, previous works considered infinite systems (both euclidean
and fractal), and studied their properties in some kind of
thermodynamic limit; typically, sending the volume to infinity
while keeping the density of reactants fixed. Therefore, they
considered the critical properties of the systems (for example,
the scaling of the density of reactants) for $t
\rightarrow\infty$, hence, for an {\it infinite} time lapse of the
reaction. One of the most important issues of this paper regards
the finite size of the systems under study. In this work we
examine explicitly finite systems where no thermodynamic limit has
to be taken. All the quantities we calculate, in particular the
total reaction time $\tau$, are hence finite, and we seek their
dependence on the finite parameters of the system (volume of the
reaction and concentration of the reactants).

In particular, in this paper we study the dynamics of a system
made up of two species particles undergoing irreversible quadratic
autocatalytic reactions according to the following scheme: $A + B
\rightarrow 2A$, with reaction probability set equal to one. All
particles move randomly and particles of different kinds react on
encounter, i.e. the reaction is strictly local and deterministic.
Autocatalytic reactions have been extensively analyzed on
Euclidean structures, both analytically and numerically
\cite{lemarchand,mai1,mai2,velikanov,warren,chaivorapoj}. A
continuous picture of this system can be attained by the Fisher
equation \cite{fisher,kolmo} which describes the system in terms
of front propagation; however, this picture will not intervene in
our calculations, that will mostly concern the low-density regime,
where a front propagation cannot be defined.

While previous works on autocatalytic reactions considered reactions
on Euclidean lattices, here we focus, as mentioned above, on low
dimensional structures ($1 < d < 2$), hence considering media whose
properties are not translationally invariant. Our investigations are
especially concerned with the role of topology in the temporal
evolution of the system. In particular, we will consider the
concentration $\rho_A(t)$ of A particles present in the system at
time $t$ and its fluctuations. From $\rho_A(t)$ it is also possible
to derive an estimate for the reaction velocity. Furthermore, we
consider the average time $\tau$ at which the system achieves its
inert state, i.e. $N_A=N$. We call this time ``Final Time''. As we
will show, $\tau$ depends on the number of particles $N$ and on the
volume $V$ of the underlying structure, meant as the total number of
sites. More precisely, it will be shown, both numerically and
analytically, that for small concentrations of the reactants the
``Final Time'' factorizes into two terms depending on $N$ and $V$
respectively. This results agrees with previous works where the
model under study was analyzed for Euclidean lattices
\cite{earlier1,earlier2}. We will also show how this dependence
could provide a practical tool for the determination of the
concentration of reactants, especially when very small
concentrations have to be detected \cite{endo}.

The plan of the paper is the following. In Sec. \ref{sec:model} we
introduce the model, we recall previous results on Euclidean
substrates (Sec. \ref{sec:model_euc}) and discuss the main
features of inhomogeneous lattices (Sec. \ref{sec:model_frac}). In
Sec. \ref{sec:Analytical} we show our analytical results
concerning those lattices; Sec. \ref{sec:NumRes} discusses the
results of numerical simulations. Sec. \ref{sec:Conclusions}
contains our conclusions.

\section{\label{sec:model}The model}
We consider a system made up of $N$ particles of two different
chemical species $A$ and $B$, diffusing and reacting on a discrete
substrate with no excluded volume effects. The volume of the
substrate is $V$; at time $t$, $N_A(t)$ and $N_B(t)$ are the number
of $A$ and $B$ particles, respectively, with $N=N_A+N_B$. We define
$\rho_A(t)=N_A(t)/V$ and $\rho_B(t)=N_B(t)/V$ as the concentrations
of the two species at time $t$.

Different species particles residing at time step $t$, on the same
node or on nearest-neighbor nodes react according to the following
mechanism:
$$
A + B \rightarrow 2A
$$
with reaction probability set equal to one, so that the process is
strictly diffusion-controlled. Notice that the previous scheme is
quite general as it also includes possible additional products
(other than $2A$) made up of some inert species of no consequences
to the overall kinetics.

The initial condition at time $t=0$ is $N_A(0)=1, N_B(0)=N-1$, with
all particles distributed randomly throughout the substrate. As a
consequence of the chemical reaction defined above, $N_A(t)$ is a
monotonic function of $t$ and, due to the finiteness of the system,
it finally reaches value $N$; at that stage the system is chemically
inert. The average time at which $N_A(t)=N$ is called ``Final Time''
and denoted by $\tau$.

The Final Time $\tau$ is of great experimental importance since it
provides an estimate of the time when reaction-induced effects
(such as side-reactions or photoemission) vanish \cite{ichimura}.

In this perspective, deviations from the theoretical prediction of
$\tau$ are, as well, noteworthy: they could reveal the existence
of competitive reactions \cite{merkin} or explain how the process
is affected by external radiation \cite{singh}.

However, one of the most interesting applications of the Final
Time is analytic \cite{endo,ishihara}: as we will show, $\tau$
sensitively depends on $N$, that is on the initial amount of
reactant. Hence, given a trace reactant, its determination can be
achieved by means of spectrophotometric measures of $\tau$.

%

Indeed, our results confirm that this technique can be extremely
sensitive \cite{endo,kato} and the determination of ultratrace
amounts is therefore possible.

We stress again that he following analysis mainly focuses on
high-diluted finite-size systems. The experimental applications
previously described concerns finite systems; furthermore, we aim to
evidence the role of the substrate topology which just emerges in
the diffusion-limited regime.

\subsection{\label{sec:model_euc}Euclidean Substrate}

The quadratic autocatalytic system for diffusing reactants on
Euclidean lattices has been analyzed in detail in earlier works
\cite{earlier1,earlier2} in the context of information spreading. It
also provides a simple model for epidemic systems: $A$ ($B$)
particles stand for (irreversibly) sick (healthy) or informed
(unaware) individuals respectively. For these systems as well a
knowledge of the rate of infection or information diffusion is of
great importance. We briefly review the results obtained in
\cite{earlier1,earlier2} for $\tau$ on Euclidean lattices.

In general, $\tau$ depends on system parameters $N$ and $L$ and, in
the low-concentration regime, and for the Euclidean lattices this
dependence can be factorized into two contributions depending on $N$
and $L$ respectively, and whose functional form depends on the
dimension of the lattice. A mean-field calculation for $\tau$
provides exact results for $d>2$ and $d=1$, while dimension $d=2$ is
marginal. Our previous results can be summarized as follows:
\begin{equation}\label{eq:Tau}
\tau(N,L) \sim \ \left\{
\begin{array}{ccc}
\displaystyle  C_1\frac{L^2}{N} & d=1 \\
\\
f_2(N)\,L^2\, \mathrm{ln}(L)\, & d=2 \\
\\
\displaystyle  C_d\frac{\gamma + \mathrm{ln}(N)}{N}\,L^d & d\geq3,
\end{array}
\right.
\end{equation}
where $C_d$ are dimension-depending constants and fitting procedures
suggest $f_2(N) = \frac{A + ln(N)}{N}$ (for further details see
\cite{earlier2}).

\subsection{\label{sec:model_frac}Inhomogeneous Structures}

\begin{figure}[tb]
\center{
\resizebox{0.70\columnwidth}{!}{\includegraphics{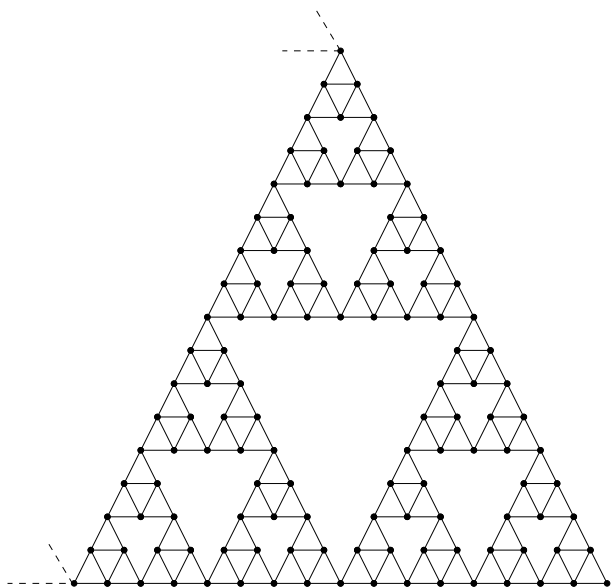}}}
\caption{\label{fig:gasket}  Sierpinski gasket of generation 4:
$V=\frac{3}{2}(3^3-1)$.}
\end{figure}

\begin{figure}[tb]
\center{
\resizebox{0.70\columnwidth}{!}{\includegraphics{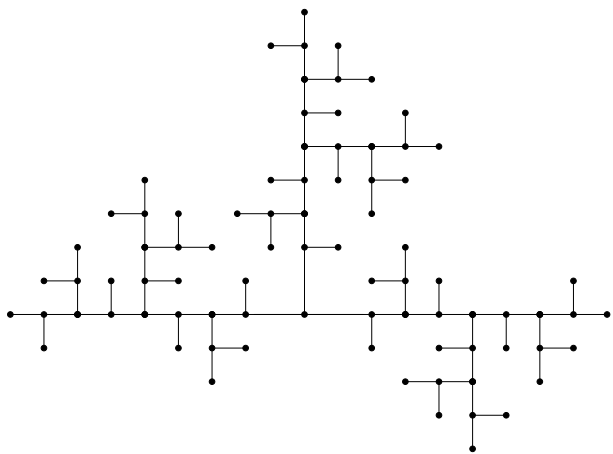}}}
\caption{\label{fig:tfractal} T-fractal of generation 4:
$V=3^4+1$.}
\end{figure}

\begin{figure}[tb]
\center{
\resizebox{0.70\columnwidth}{!}{\includegraphics{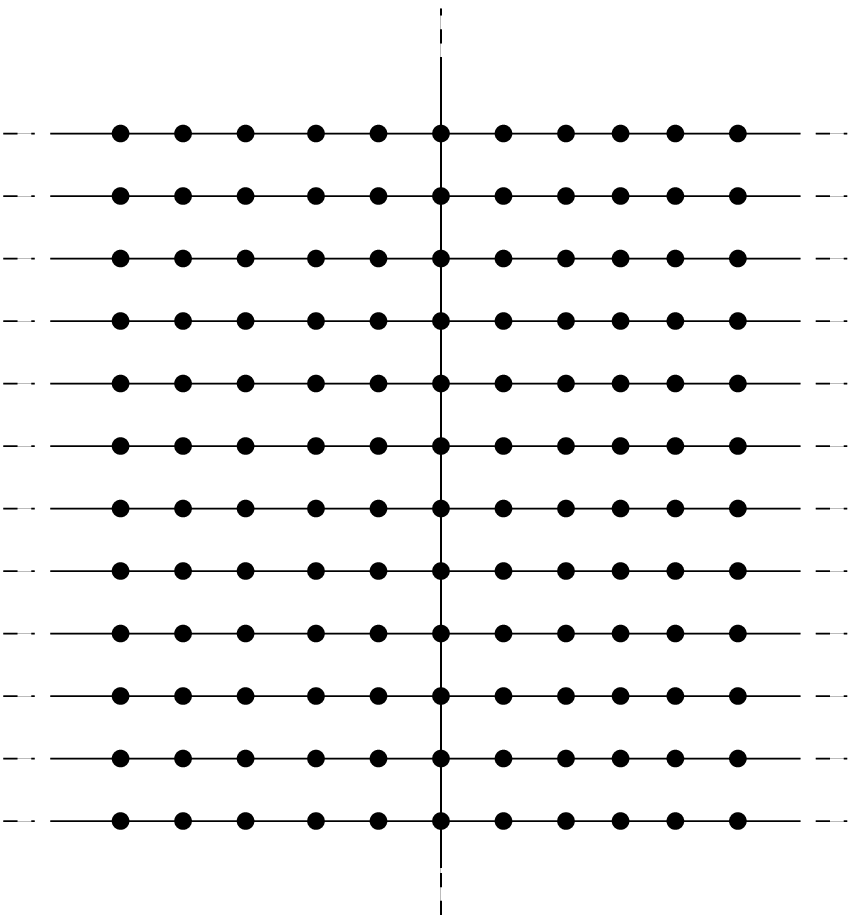}}}
\caption{\label{fig:Comb}  Comb Lattice}
\end{figure}

A number of problems in physics and chemistry are related to random
walks on homogeneous structures. In the last few years there is a
growing interest in the theory of diffusion in low and fractal
dimensions \cite{havlin}. In fact, most of the materials existing in
nature show a disordered, non-crystalline geometrical structure.
Indeed, fractals are good models for disordered systems since they
usually display a dilation symmetry which is a fractal
characteristic. Apart from the applications point of view, low
dimensional systems are of great theoretical importance. As
previously underlined, when diffusion is considered, low dimensional
structures dramatically affect the kinetic laws.

Fractals are self-similar structures exhibiting dilation symmetry.
Differently from Euclidean structures, their description requires at
least two (typically noninteger) different dimensions. The first is
the fractal dimension $d_f$, that gives the dependence of the volume
of the system (i.e., the number of sites it contains) on its linear
size
$$
V(L)\sim L^{d_f}.
$$
The second is the spectral dimension $\tilde{d}$, which governs
(among other phenomena) the long-time properties of diffusion on the
lattice. Indeed, if we consider a random walkers starting from a
given site $i$ of the lattice, the probability $P_{ii}(t)$ of
returning back to the starting point at time $t$, at large times,
follows the law
$$
P_{ii}(t)\sim t^{-\tilde{d}/2}.
$$
Also, the number $S(t)$ of different sites visited by the random
walker at time $t$ is for large times:
\begin{eqnarray}\label{eq:differentSites}
   & S(t)\sim t^{\tilde{d}/2} &\;\mathrm{for}\;\tilde{d}\leq 2 \nonumber\\
   & S(t)\sim t & \;\mathrm{for}\;\tilde{d} > 2
\end{eqnarray}
For $\tilde{d}<2$ the random walker returns to its starting site
with probability 1 and the lattice is called {\it recurrent}; for
$\tilde{d}>2$ the probability of return is less than 1 and the
lattice is called {\it transient} (lattices with $\tilde{d}=2$ have
to be discussed case by case). For $\tilde{d}<2$ the walker is also
said to perform a {\it compact exploration} of the space
\cite{degennes}, since the (fractal) dimension of the random walk
trajectory is greater than the dimension $d_f$ of the underlying
lattice.

The spectral dimension is of a more general interest than the
fractal dimension, since it can also be defined for inhomogeneous
structures that lack a dilation invariance, hence for which a
fractal dimension $d_f$ cannot be defined.

The two fractals we consider in this paper, the Sierpinski gasket
and the T-fractal, have both $\tilde{d}<2$.

The Sierpinski gasket is generated by iterating in a recurrent
fashion a generating cell consisting of a triangle
(Fig.\ref{fig:gasket}). The number of iterations is called the
generation $g$ of the fractal. The total number of triangles after
$g$ iterations is $3^g$, while the total number of sites (hereafter
called volume $V$) is $V=3 \frac{3^g + 1}{2}.$ The linear size of
the gasket is given by $2^g$. The Sierpinski gasket has fractal
dimension $d_f=\frac{\log3}{\log2}\approx1.584$ and spectral
dimension $\tilde{d}=\frac{\log 9}{\log 5}\approx1.365.$

The T-fractal is constructed from a 4-sites T-shaped generating cell
(Fig.\ref{fig:tfractal}). It has fractal dimension
$d_f=\frac{\log3}{\log2}\approx1.584$ and spectral dimension
$\tilde{d}=\frac{\log 9}{\log 6}\approx1.226.$

The third structure we will consider, the comb lattice
(Fig.\ref{fig:Comb}), does not present dilation invariance; hence,
the $d_f$ cannot be defined. As we said above, it is still possible
to define a spectral dimension that turns out to be $\tilde{d}=3/2$.

Some linear problems have already been solved exactly on these
structures using renormalization groups technique (see for example
\cite{broek}). However, for many interacting diffusing particles an
exact solution is not feasible and we rely mainly on numerical
simulations.

\section{\label{sec:Analytical} Analytical Results}

In this section we study the irreversible autocatalytic reaction
occurring in a close system by means of a mean field approximation.
This kind of approach is very different from those previously
adopted for this kind of system.

An analytical result for the dependence of the Final Time $\tau$ on
$V$ and $N$ is difficult to find: approximate calculations can be
carried out in the two limit regimes of high and low concentration.

For high concentration ($\rho\gg 1$) the results found in Euclidean
structures \cite{earlier1} continue to hold for inhomogeneous
geometries: the $A$ particles occupy a {\it connected} region of the
space for all $t$. The border of this region expands at a fixed
velocity, such that at time $t$ the region covers all the sites
whose chemical distance from the starting point of the seed particle
is $\leq 2\,t$. Hence, for a finite system like those we are
considering here, the Final Time is $\tau=l_{\mathrm{max}}/2$, where
$l_{\mathrm{max}}$ is the chemical distance of the most distant
point on the lattice, starting from the seed particle. On Euclidean
geometries this yields $\tau=L/2$ for $d=1$ and $\tau=L$ for $d\geq
2$. On the other hand, on inhomogeneous structures the dependence on
$L$ is not so simple, since it involves taking the average with
respect to all possible starting points for the seed particle;
anyway, the crossover between this high-concentration regime and the
low-concentration one remains apparent.

Our mean-field approach is based on a different point of view and
the assumptions introduced make it valid just in the
low-concentration regime. In this approach we focus on collective
quantities lacking the spatial dependence.

In particular our hypothesis is that the time elapsing between a
reaction and the successive one is long enough that the spatial
distribution of reactants can be considered as uniform. This
assumption corresponds to a mean field approach since we neglect
correlations between spatial position of reactants; in other words
we neglect the fact that for small times after a reaction the two A
particles are likely to be find nearby. This kind of hypothesis is
therefore valid for small concentration of reactants, i.e. $\rho \ll
1.$ As a consequence of this hypothesis we can just focus on
two-body interactions among particles since the event of three or
more reactants interacting together is unlikely at small densities.

First of all we consider the final time $\tau$. Let us define
$\langle t_n \rangle$ the time elapsing between the $n-1$-th first
encounter among different particles and the $n$-th one. This time
corresponds to the average time during which there are just
$N_A(t)=n$ particles in the systems. In the mean field approximation
this is proportional to the trapping time in the presence of $n$
traps randomly distributed through a volume $V$. For compact
exploration of the space ($\tilde{d}<2$) \cite{degennes}, the
average trapping time $t_{\mathrm{trap}}$ for a random walker in a
distribution of $N-N_A$ randomly distributed moving traps is given
by \cite{oshanin}

$$
t_{\mathrm{trap}}\sim\rho_{\mathrm{trap}}^{-2/\tilde{d}}=\left(\frac{V}{N_A(t)}\right)
^{2/\tilde{d}},
$$
since the density of traps is $\rho_{traps}=N_A(t)/V$. Here, the
symbol $\sim$ denotes proportionality.

This is the trapping time for one particle in a background of moving
traps; we are interested in the average trapping time of the first
out of $N-N_A$ particles, that for rare events is just the same time
rescaled by a factor $N-N_A$ (the number of $B$ particles):

\begin{equation}
\label{eq:Tau_step} \langle t_n \rangle = V^{2 / \tilde{d}}
\frac{N_A^{-2/\tilde{d}}}{N-N_A}.
\end{equation}

The time $\tau$ can therefore be written as
\begin{equation}
\label{eq:Tau_sum} \tau = \displaystyle \sum_{N_A=1}^{N-1} \langle
t_{N_A} \rangle.
\end{equation}
For this sum there exists no closed form; however, in the limit $N
\rightarrow \infty$ we can adopt a continuous approximation and
replace the sum with an integral. We then find
\begin{equation}
\label{eq:Tau_mf} \tau \sim V^{2/\tilde{d}} \left[
\frac{\tilde{d}}{(2-\tilde{d})N} + N^{-2/\tilde{d}} (\log N +
H_{2/\tilde{d}}) + \mathcal{O}(N^{-1}) \right ],
\end{equation}
where $H_m$ is the harmonic number
$$
H_m=\sum_{k=1}^{m}\frac{1}{k}.
$$
In particular, the leading-order contribution for a one-dimensional
system ($\tilde{d}=d=1$) is
\begin{equation}
\label{eq:Tau_mf_1d} \tau \sim \frac{V^2}{N}
\end{equation}
and for a two-dimensional lattice ($\tilde{d}=d=2$)
\begin{equation}
\label{eq:Tau_mf_2d} \tau \sim V\frac{\log N+\gamma}{N}
\end{equation}
where $\gamma$ is the Euler-Mascheroni constant. For
$1<\tilde{d}<2$, the expression in (\ref{eq:Tau_mf}) interpolates
between (\ref{eq:Tau_mf_1d}) and (\ref{eq:Tau_mf_2d}).

\begin{figure}[!t]
\center{
\resizebox{0.70\columnwidth}{!}{\includegraphics{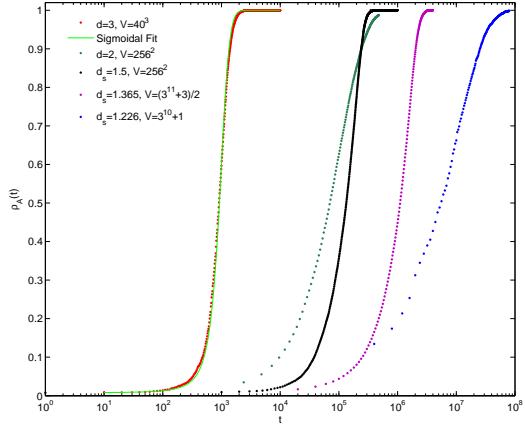}}}
\caption{\label{fig:sigmoidal_compare}  (Color online)
Concentration of $A$ particles $\rho_A(t)$ vs time $t$ for a
system made up of $N=128$ particles embedded on different
structures, as shown by the legend. The best fit for the cubic
lattice is also shown and it is given by a pure sigmoidal.}
\end{figure}

The mean-field extension just performed also allows to derive some
insights into the temporal behavior displayed by $N_A(t)$. In fact,
being $\tau_n$ the average time at which $N_A=n$, from Eq.
\ref{eq:Tau_sum} we can write
\begin{equation}
\label{eq:Tau_sum_n} \tau_n = \displaystyle \sum_{N_A=1}^{n-1}
\langle t_{N_A} \rangle = f(n).
\end{equation}
Now, we estimate $N_A(t)$ as $$N_A(t)=f^{-1}(t)$$ whose numerical
solution provides an S-shaped curve to be compared (see
Fig.~\ref{fig:sigmoidal_compare}) with the sigmoidal curve obtained
from a standard mean-field approximation \cite{earlier1}
\begin{equation}
\label{eq:sigmoide} N_A(t)=\frac{N}{(N-1)e^{-Npt}+1},
\end{equation}
where $p$ is a quantity proportional to the concentration $\rho$
that in practice must be adjusted within the fitting procedure.

\section{\label{sec:NumRes}Simulations}

In this section we show results obtained with numerical
simulations performed on the Sierpinski gasket, on the T graph and
on comb structures.

\begin{figure}[t]\center{
\resizebox{0.45\columnwidth}{!}{\includegraphics{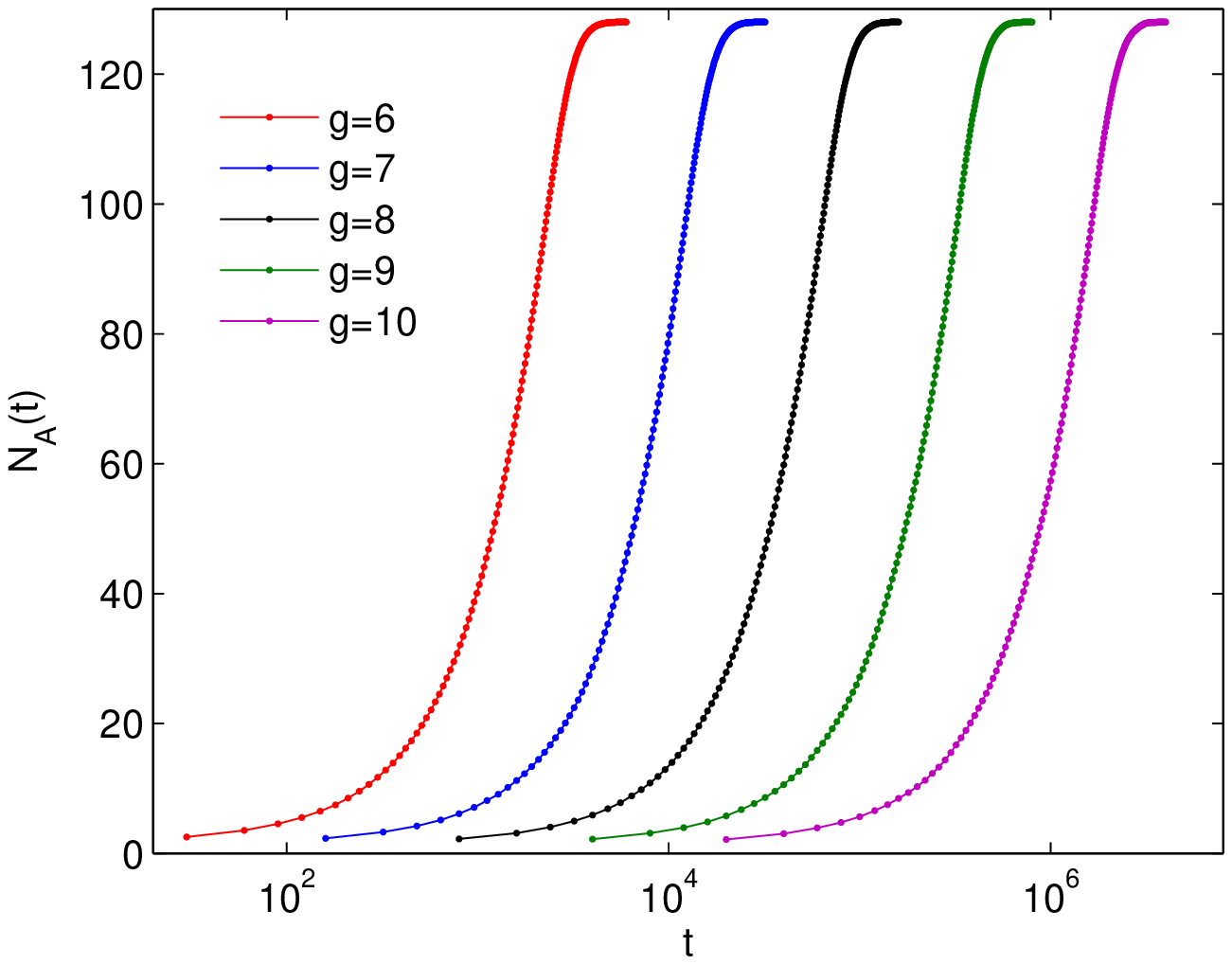}}
\resizebox{0.45\columnwidth}{!}{\includegraphics{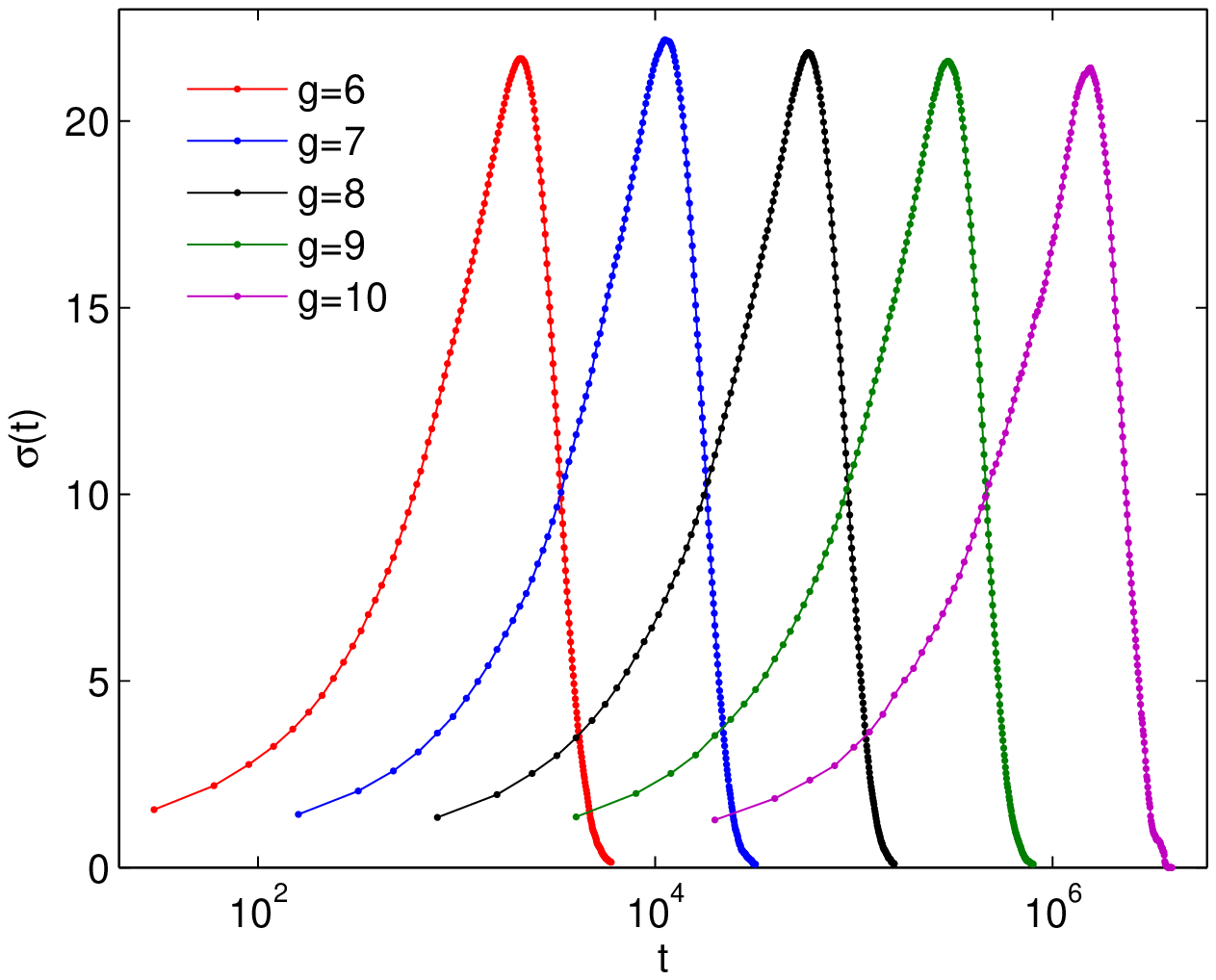}}}
\caption{\label{fig:Distr_S}(Color online) Data depicted in the
figures above refers to a system of $N=128$ particles diffusing on
a Sierpinski gasket of five different generations. Each
generations is depicted in different colors, as shown by the
legend. Left: Number of $A$ particles $N_A(t)$ present in the
system vs time $t$. Right: Fluctuations $\sigma(t) = \sqrt{\langle
N_A^2\rangle - \langle N_A \rangle^2}$ as a function of time}
\end{figure}

First of all we consider the dependence on $t$ displayed by
$N_A(t)$. In Fig.~\ref{fig:sigmoidal_compare} and
\ref{fig:Distr_S} we show data obtained for the Sierpinski gasket.
In particular, in the latter figure we also provide a comparison
with results obtained for the T-graph, the comb lattice, the
square lattice and the cubic lattice. Consistently with results
found in \cite{earlier1}, for transient lattices $N_A(t)$ is well
fitted by the sigmoidal function of Eq. \ref{eq:sigmoide}, while
for low-dimensional structures deviations are expected.

In fig.\ref{fig:sigmoidal_compare} we show the time evolution
$N_A(t)$ for substrates with the same total number of particles
$N$ and with (approximately) the same volume $V$, but different
spectral dimension $d$. In order of decreasing $d$, they are: the
cubic lattice ($\tilde{d}=d=3$), the square lattice
($\tilde{d}=d=2$), the comb graph ($\tilde{d}=3/2$), the
Sierpinski gasket fractal ($\tilde{d}=\frac{\log 9}{\log
5}\approx1.365$) and the T-fractal ($\tilde{d}=\frac{\log 9}{\log
6}\approx1.226$). We remind that the spectral dimension describes
the long-range connectivity structure of the substrate and the
long-time diffusive behavior of a random walker on the substrate.
In particular, from Eq.~\ref{eq:differentSites}, we expect that
for substrates $\tilde{d}\leq 2$ the number of different sites
visited by each walker will grow faster as $\tilde{d}$ increases,
and so will the number of meetings between walkers. Hence, we
expect the curves $N_A(t)$ to grow faster, and saturate earlier,
with increasing $\tilde{d}$ ($N$ and $V$ being fixed). This is
precisely what happens, as shown in
Fig.~\ref{fig:sigmoidal_compare} (except for the saturation time
$\tau$ of the comb lattice, which will be discussed below). For
$\tilde{d}\geq 2$ (e.g., $\tilde{d}=3$ in the figure), $N_A(t)$ is
independent of $\tilde{d}$ and is fitted by a pure sigmoidal
function.

\begin{figure}[t]\center{
\resizebox{0.70\columnwidth}{!}{\includegraphics{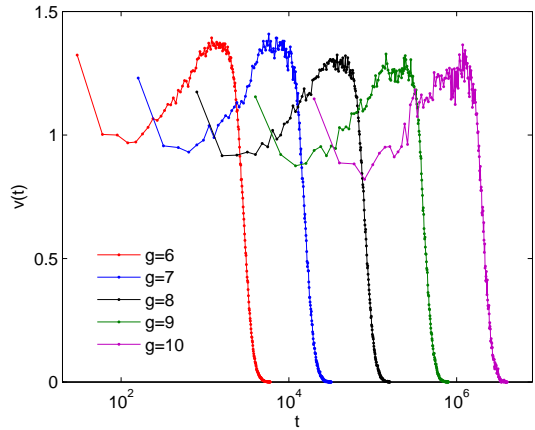}}}
\caption{\label{fig:Velocity}  (Color online) Reaction velocity
$v(t)$ for a system of $N=128$ particles diffusing on a Sierpinski
gasket; five different generations (depicted in different colors)
are shown. The reaction velocity is defined in
Eq.~\ref{eq:velocity}}
\end{figure}

From $N_A(t)$ one can derive the rate of reaction
\begin{equation}
\label{eq:velocity} v(t)=\partial_t N_A(t).
\end{equation}
As you can see from the numerical results in
Fig.~\ref{fig:Velocity}, in agreement with the theoretical one,
$v(t)$ is an asymmetrical curve exhibiting a maximum at a time
denoted by $t_v$. This time obviously corresponds to a flex in
$N_A(t)$ which scales with the volume of the structure according
to the following:
$$
t_v \sim V^{2/\tilde{d}.}
$$
This is the same dependence shown by $\tau$ (see below), and
corresponds to a situation in which the population of the two
species are about the same ($N_A=N_B=N/2$). Analogous results can be
obtained for the T-fractal.

Furthermore, the profile shown in Fig.~\ref{fig:Velocity} also
suggests that the efficiency of the autocatalytic reaction is not
constant in time but, provided the number $N$ of particles is
conserved, it exhibits a maximum when the number of $B$ particles is
about $N/2$.

A similar result may be derived for the variance $\sigma_A(t)$ of
the number of $A$ particles present on the substrate.

Interestingly, fluctuations display a maximum at a time
$t_{\sigma}$ which, again, depends on the system size with the
same law as $\tau$. Notice that $t_{\sigma} > t_v$ and
$N_A(t_v)=\frac{N}{2}.$

Finally, we consider the time $\tau$ representing the average time
at which the autocatalytic reaction stops since all $B$ particles
have been transformed into $A$ particles. In general this quantity
depends on system parameters $V$ and $N=N_A + N_B$ and, as we will
show, its functional form is significantly affected by the topology
of the substrate.

\begin{figure}[t]\center{
\resizebox{0.70\columnwidth}{!}{\includegraphics{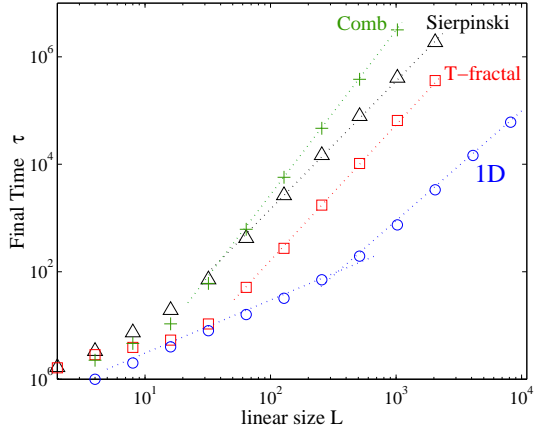}}}
\caption{\label{fig:scaling_tau}  Scaling of $\tau$ with the
linear size of the system for a 1-dimensional chain (blue
circles), a Sierpinski gasket (black triangles), a T-fractal (red
squares), and a comb lattice (green plus signs) on a
double-logarithmic scale. The number of reactants is fixed at
$N=1024$ for all systems. Dotted lines highlight the
low-concentration regime ($L\gg 1$), corresponding to a power law
for all systems. For the 1-dimensional chain, the linear
high-concentration regime (low $L$) is also highlighted.}
\end{figure}

\begin{figure}[b]
\center{
\resizebox{0.45\columnwidth}{!}{\includegraphics{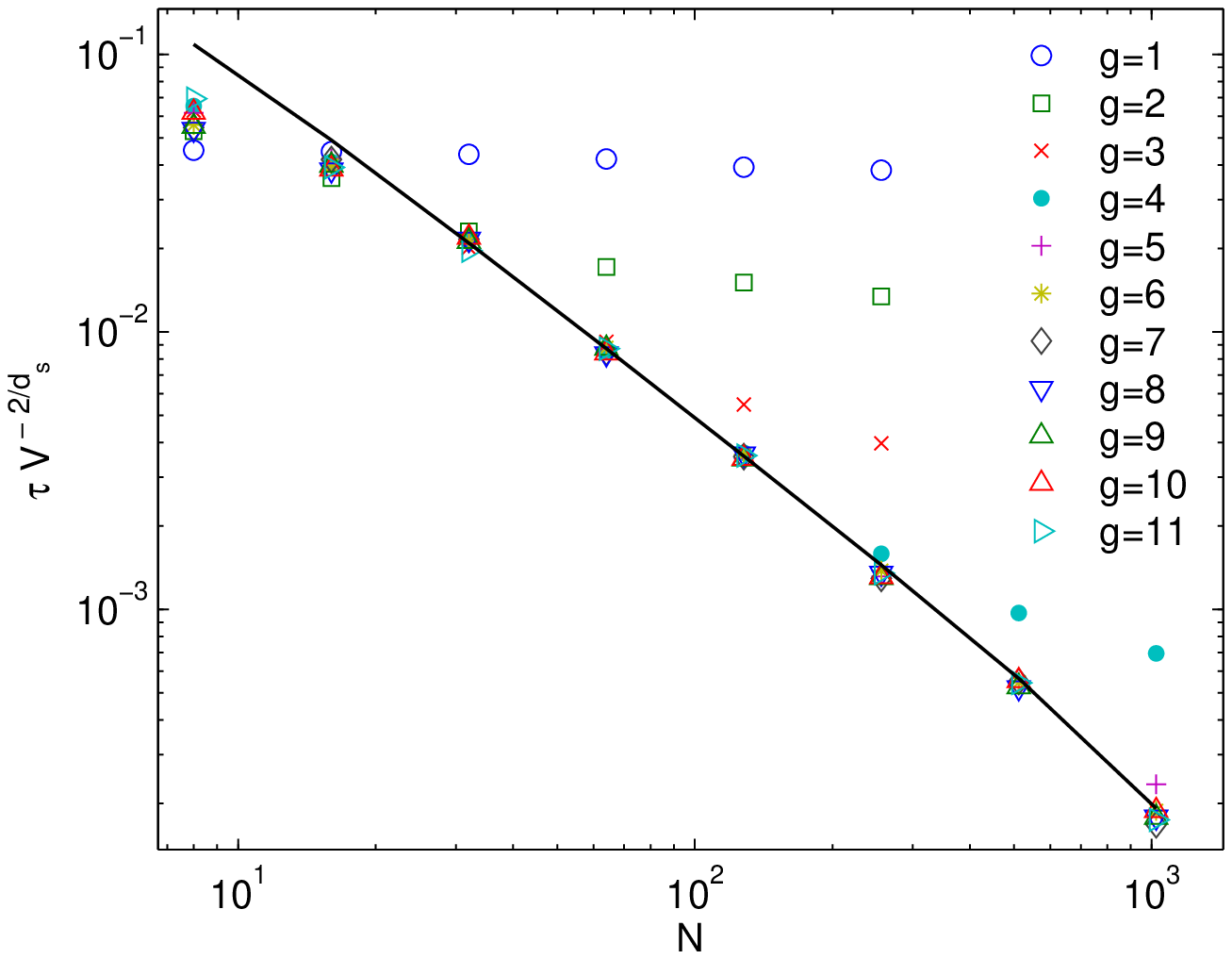}}
\resizebox{0.45\columnwidth}{!}{\includegraphics{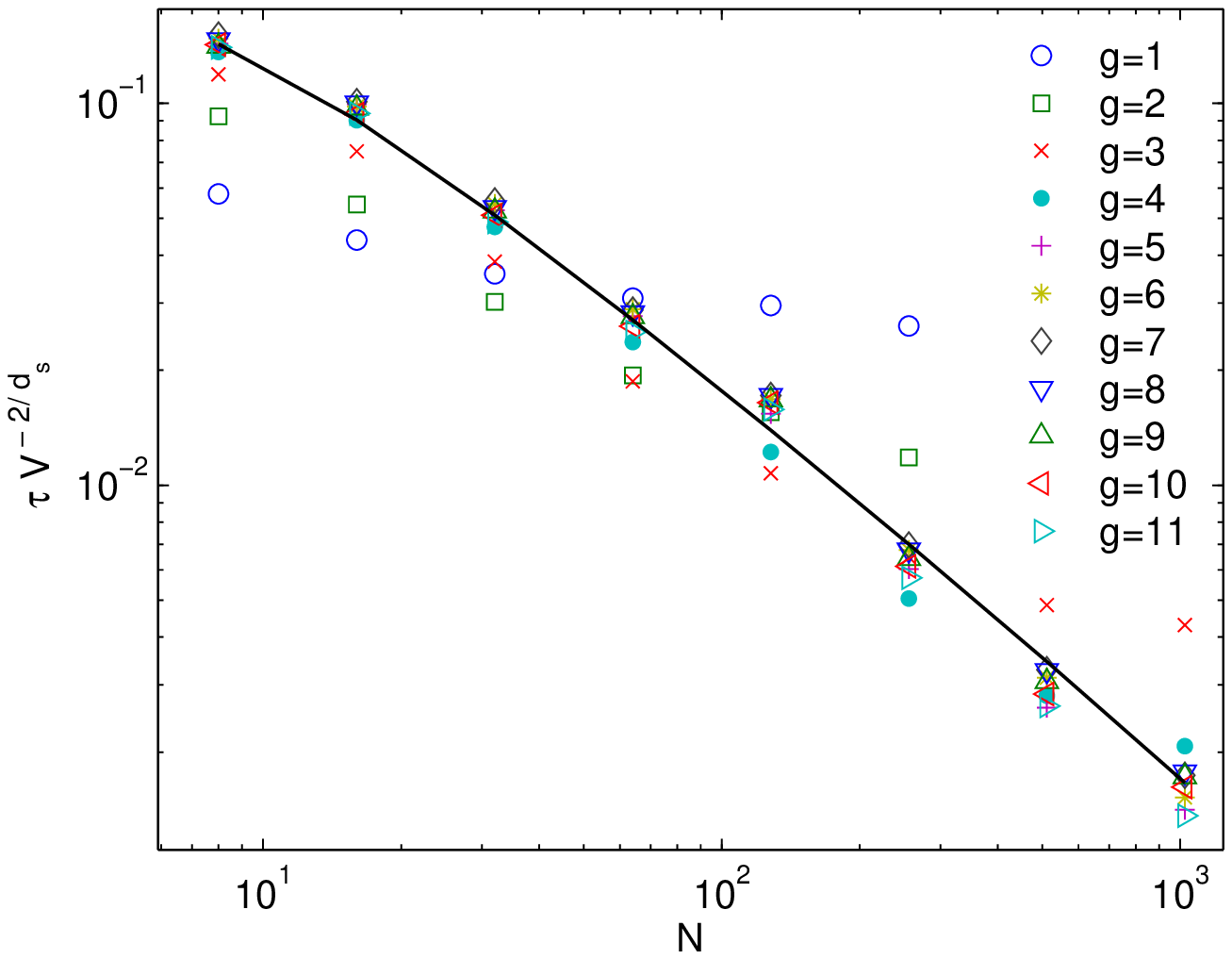}} }
\caption{\label{fig:Tau}  (Color online) Rescaled Final time $\tau
V^{-2/ \tilde{d}}$ vs number of particles $N$ for Sierpinski
gasket (top) and T-fractal (bottom). Different symbols and colors
distinguish different generation as explained by the legend. The
line provides the best fit in agreement with equation
\ref{eq:Tau_rewritten}.}
\end{figure}

The dependence of $\tau$ on the system size $L$ (or the volume
$V=L^{d_f}$) clearly displays two different regimes, as shown in
Fig. \ref{fig:scaling_tau}. In both cases, $\tau$ increases with
$L$, but in the high-concentration regime the growth is less
rapid; in particular (as shown for $d=1$ in the figure), it is
proportional to $L$ for Euclidean lattices.

In the low-concentration regime, where we can assume that reactions
only occur among two particles, the mean-field-like calculation
explained in the previous section holds and we expect $\tau$ to vary
with $N$ and $V$ according to Eq. \ref{eq:Tau_mf} which can be
rewritten as:
\begin{equation}
\label{eq:Tau_rewritten} \tau V^{-2 / \tilde{d}} =  [ \frac{1}{N} +
N^{-2 / \tilde{d}} (\log N - H_{\frac{2}{\tilde{d}}}) ].
\end{equation}
Hence in Figure \ref{fig:Tau} we plotted $\tau V^{-2 / \tilde{d}}$
vs $N$ and we fitted data according with the r.h.s . of the previous
equation.

It can be seen that for small densities all the data collapse.
Moreover, in that region, the fit coefficients introduced are in
good agreement with theoretical predictions.

It should be underlined that the average final time depends
non-trivially, on $N$ and $V$, viz. $\tau$ does not depend directly
on the total concentration $\rho$, though the dependence on $N$ and
$V$ can be factorized.

The agreement of formula \ref{eq:Tau_rewritten} for the comb lattice
is less good. In particular, it seems that the dependence of $\tau$
on $N$ and $V$ can still be factorized, but that the exponent for
$V$ is rather $8/3$ than $2/\tilde{d}=4/3$. This may mean that the
particular mean field approximation we have made does not hold
anymore for strongly inhomogeneous structures such as combs; this
point is still under investigation.
\begin{figure}[b]
\center{\resizebox{0.70\columnwidth}{!}{\includegraphics{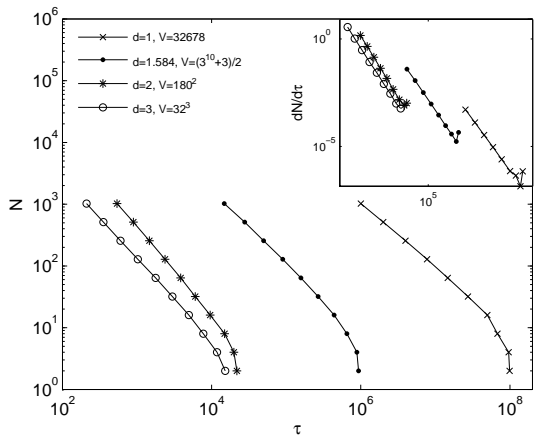}}}
\caption{\label{fig:N-Tau} Loglog scale plot of the reactant
amount $N$ vs Final time $\tau$; as shown in the legend, different
substrate structures (with approximately the same volume) have
been compared. In the inset the derivative $\frac{dN}{d\tau}$ is
depicted again as a function of $\tau$. Lines are guides for the
eyes.}
\end{figure}

As explained in Section \ref{sec:model}, experimental measures of
$\tau$ are useful in monitoring trace reactants. Indeed, our results
show that $\tau=f_{\tilde{d}}(N)g_{\tilde{d}}(V)$ and therefore,
once the substrate size is fixed, the initial amount of reactant can
be expressed as $N =
f_{\tilde{d}}^{-1}(\frac{\tau}{g_{\tilde{d}}(V)})$.

A proper estimate of the sensitivity of this method is provided by
the derivative $\frac{dN}{d\tau}$: the smaller the derivative and
the larger the sensitivity. In Fig.~\ref{fig:N-Tau} we depicted
numerical results for both $N$ and its derivative
$\frac{dN}{d\tau}$ as a function of $\tau$; topologically
different substrates are also compared, all sharing,
approximately, the same volume. The numerical plots provided
allows a qualitative analysis and comparison among the different
structures considered; a more quantitative inspection can be
outlined after a proper calibration procedure.

First of all, notice that the characteristic curves are well defined
and they allows a univocal determination of $N$ from $\tau$.
Moreover, the sensitivity of this analytic technique is better for
small values of $N$ (with $N>2$) and, interestingly, for
low-dimensional substrates. Indeed, once $V$ is fixed, when $d \leq
2$, the technique sensitivity is improved by lowering the substrate
dimension. On the other hand, when $d>2$, the resulting curves are
overlapped, hence no improvement is achieved. It should be
underlined that the high sensitivity attained just for small
concentrations of reactants makes this analytic technique very
suitable for the determination of ultratrace amounts of reactants,
which is of great experimental importance \cite{rose,priest,veer}

\section{\label{sec:Conclusions} Conclusions}

We have presented a model that considers the autocatalytic reaction
$A+B\rightarrow 2\,A$ on non-Euclidean, low-dimensional
($\tilde{d}<2$), finite-size substrates, characterized by a volume
$V$ and a total number of reacting particles $N$.

We showed by analytical calculations that the Final Time $\tau$ (the
total time span of the reaction) displays two different regimes, for
high and low concentrations, with a different dependence on $V$ and
$N$. In particular, the functional law for low concentrations can be
recovered by means of a mean-field approximation. In fact, with
respect to the standard one, our mean-field approach is able to take
into account the topological effect arising from a low-dimensional
substrate.

Numerical simulations corroborated these results for fractals,
while simulations on strongly inhomogeneous lattices (combs) hint
at a quantitatively different behavior.

Theoretical results concerning the average Final Time find important
applications in analytical fields, where measures of $\tau$ are
exploited for detecting trace reactants. Our results suggest that
the sensitivity of such technique is affected not only by the
reactant concentration, but also by the topology of the structure
underlying diffusion. More precisely, a small concentration of
reactants implies a better sensitivity, hence allowing the
determination of ultra-trace reactants. Moreover, at the same
concentration, and for low-dimensional ($d<2$) substrates, by
reducing the dimension $d$ the sensitivity can be further improved.

%
%

\end{document}